\documentclass[]{spie}  

\usepackage{amsmath,amsfonts,amssymb}
\usepackage{graphicx}
\usepackage[colorlinks=true, allcolors=blue]{hyperref}

\title{The Cherenkov Telescope Array On-Site integral sensitivity: observing the Crab}

\author[a]{Valentina Fioretti}
\author[a]{Andrea Bulgarelli}
\author[b]{Fabian Sch\"ussler}
\author[c]{for the CTA Consortium}
\affil[a]{INAF Istituto di Astrofisica Spaziale e Fisica Cosmica Bologna, Via P. Gobetti 101, 40129, Bologna, Italy}
\affil[b]{Irfu / CEA-Saclay, 91191 Gif-sur-Yvette, France}
\affil[c]{Full consortium author list at http://cta-observatory.org}

\authorinfo{Further author information: (Send correspondence to V. Fioretti)\\E-mail: fioretti@iasfbo.inaf.it, Telephone: +39 051 6398772}

\begin{document} 
\maketitle

\begin{abstract}
The Cherenkov Telescope Array (CTA) is the future large observatory in the very high energy (VHE) domain. Operating from 20 GeV to 300 TeV, it will be composed of tens of Imaging Air Cherenkov Telescopes (IACTs) displaced in a large area of a few square kilometers in both the southern and northern hemispheres. 
Thanks to the wide energy coverage and the tremendous boost in effective area (10 times better than the current IACTs), for the first time a VHE observatory will be able to detect transient phenomena in short exposures. The CTA/DATA On-­Site Analysis (OSA) is the system devoted to the development of dedicated pipelines and algorithms to be used at the CTA site for the reconstruction, data quality monitoring, science monitoring and real­time science alerting during observations. 
The minimum exposure required to issue a science alert is not a general requirement of the observatory but is a function of the astrophysical object under study, because the ability to detect a given source is determined by the integral sensitivity which, in addition to the CTA Monte Carlo simulations, providing the energy-dependent instrument response (e.g. the effective area and the background rate), requires the spectral distribution of the science target. 
The OSA integral sensitivity is computed here for the most studied source at Gamma-rays, the Crab Nebula, for a set of exposures ranging from 1000 seconds to 50 hours, using the full CTA Southern array. The reason for the Crab Nebula selection as the first example of OSA integral sensitivity is twofold: (i) this source is characterized by a broad spectrum covering the entire CTA energy range; (ii) it represents, at the time of writing, the standard candle in VHE and it is often used as unit for the IACTs sensitivity.
The effect of different Crab Nebula emission models on the CTA integral sensitivity is evaluated, to emphasize the need for representative spectra of the CTA science targets in the evaluation of the OSA use cases. Using the most complete model as input to the OSA integral sensitivity, we obtain a significant detection of the Crab nebula (about 10\% of flux) even for a 1000 second exposure, for an energy threshold less than 10 TeV.
\end{abstract}


\section{INTRODUCTION}
\label{sec:intro}  
Rapid and unpredictable short term variability, with timescales from milliseconds to hours, is a common feature of many High Energy (HE) and Very High Energy (VHE) sources. Current observations by space-based instruments Fermi/LAT [\citenum{2009ApJ...697.1071A}] and AGILE/GRID [\citenum{2009A&A...502..995T}], operating up to the GeV domain, have resulted in hundreds of Astronomy Telegrams (ATels) alerting the astrophysical community on flaring emission, with $<$ hr timescale variability, by Gamma-ray Bursts (GRBs), Gamma-ray binaries, novae, and many AGNs. The combination of good sensitivity for short exposures and a fast reaction system are the key to enable multiwavelength and simultaneous observation campaigns and to reveal the physics processes and the particle populations behind such violent and energetic phenomena.
In the VHE band, current Imaging Air Cherenkov Telescopes (IACTs) have only been able to detect short (up to minute timescale) variability in a handful of extreme blazar objects (e.g. PKS 2155-304, Mrk 501, Mrk 421, PKS 1222+21). The next generation of IACT systems, the Cherenkov Telescope Array (CTA [\citenum{2011ExA....32..193A}]), will cover an unprecedented effective area (more than 10$^{6}$ m$^{2}$) on a wide energy range, from 20 GeV to 300 GeV, by means of tens of IACTs in the Southern and Northern hemispheres.
A differential sensitivity more than 10 times better than current IACTs is foreseen along the entire energy range. The basic idea behind the CTA design is to build three different classes of telescopes: few large-sized (23 m diameter) telescopes (LSTs) to image the showers at low energies, a few tens of medium-sized telescopes (MSTs), to ensure a good background rejection and survey capabilities in the TeV band, and many ($\sim70$) small-sized telescopes (SSTs) - currently to be built in the Southern site only - to maximise the event rate at the highest energies.
For flaring events below 1 hr timescale,  CTA will be orders of magnitude more sensitive than Fermi/LAT in the overlapping energy range  (25-75 GeV [\citenum{2013APh....43..348F}]). The ability to perform the very first survey of short-term variability transients at VHE will be coupled with the ability to quickly react to such events for rapid re-pointings, follow-ups, and communication to/from external multiwavelength facilities. The DATA [\citenum{2015arXiv150901012L}] On-Site Analysis (OSA [\citenum{2015arXiv150901963B}]) is the system to be run at the CTA site for the reconstruction, data quality monitoring, science monitoring and real-time science alerting during observations.
In terms of science monitoring and alerting, the CTA/OSA is composed by:
\begin{itemize}
\item the Real-Time Analysis (RTA) or Level A analysis [\citenum{2013arXiv1307.6489B}]: the analysis pipeline must issue science alerts within 30 seconds from the last acquired event, achieving an integral sensitivity not worse by more than a factor of 3 than the one of the final analysis (for the detection of a $\rm dN/dE\propto E^{−2.5}$ spectrum source on a timescale of one minute);
\item the Level-B analysis: the analysis takes place at the end of the observation, or the morning after, with a minimum differential sensitivity two times worse than the nominal one with a latency of 10 hours.
\end{itemize}
Since the integral sensitivity, i.e. the minimum detectable flux above a certain energy, depends on the spectral distribution of the source used as reference target, RTA performance can vary not only according to the subarray configuration that is performing the observation, but also to the class of transient that is detecting.
This paper is focused on the evaluation of the integral sensitivity when observing a Crab-like source at short exposures. A particular focus is dedicated to the impact of different models and/or approximations of the spectral energy distribution of the Crab Nebula in the evaluation of the integral sensitivity. 
Our results represent the first step in the general feasibility study of the OSA pipeline, with the final aim of selecting the minimum exposure, for each CTA science target, required to achieve a significant detection.  PROD2 Monte Carlo simulations for the Southern site, composed by 100 telescopes, are used throughout the analysis (see [\citenum{2013APh....43..171B}] for a review of CTA MC simulations).

\section{THE VERY HIGH ENERGY SPECTRUM OF THE CRAB NEBULA}\label{sec:crab}
\begin{figure} [b]
   \begin{center}
   \begin{tabular}{c} 
   \includegraphics[width=0.5\textwidth]{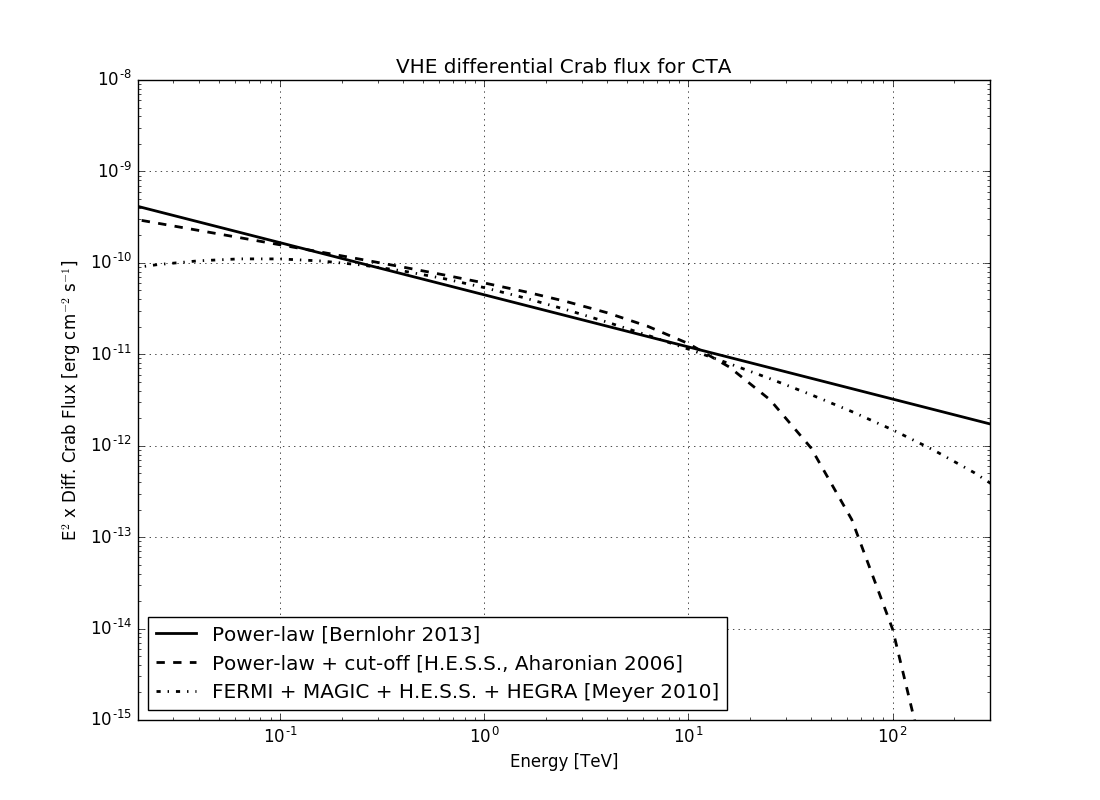}
   \includegraphics[width=0.5\textwidth]{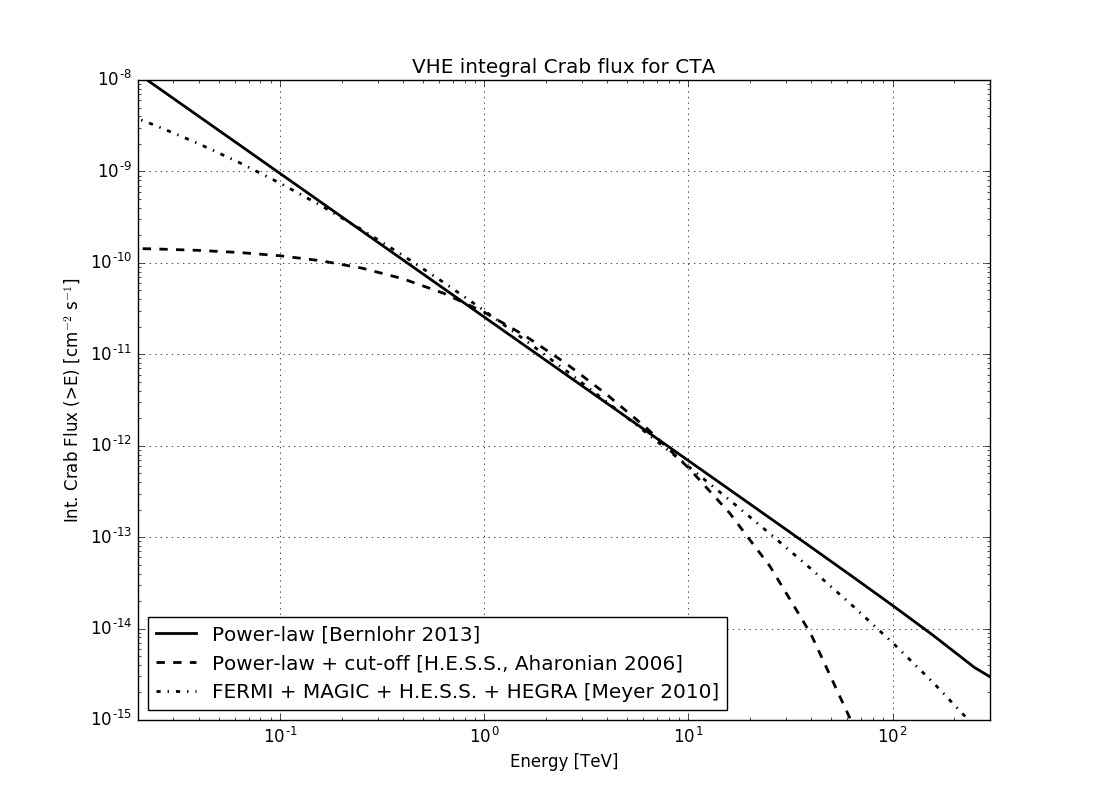}
   \end{tabular}
   \end{center}
   \caption 
   { \label{fig:model}Differential (left panel) and integral (right panel) flux of the Crab nebula in the 20 GeV - 300 TeV energy range for three models: a simple power-law (solid line), a power-law with an exponential cut-off (dashed line), and the numerical cross-correlation (dashed-dot line).}
   \end{figure}

Located at (l,b) = (184.6, -5.8) in the Galactic plane at a distance $\sim2$ kpc [\citenum{1968AJ.....73..535T}], the Crab Nebula, the supernova remnant resulting from a core-collapse in 1054 AD, is one of the most studied objects of our Universe. Observations have been carried out at all wavelengths, from radio to VHE gamma-rays. In the generally accepted picture, the pulsar powered wind nebula emits VHE photons by inverse Compton (IC) scattering of high energy electrons from the outflowing pulsar wind (see [\citenum{2010ApJ...708.1254A}] and references therein).
Despite the latest observations from H.E.S.S. [\citenum{2006A&A...457..899A}] and Fermi/LAT [\citenum{2010ApJ...708.1254A}], the current sensitivity available above 1 GeV is not enough to clarify the exact model behind the seed IC photon population and the magnetic fields into play. In addition, the unexpected flaring behavior discovered by AGILE/GRID and Fermi/LAT above 100 MeV in 2010 [\citenum{2011Sci...331..736T}] is yet to be confirmed at higher energies.
\\
In the present work, however, we are only interested on the quiescient Crab Nebula high energy spectral distribution to be used as the prototype PWN for our first RTA science case. We choose the Crab Nebula not only because of its broad spectrum but more importantly because it represents - at the time of writing - the standard candle in VHE [\citenum{2010A&A...523A...2M}]. 
\\
For science targets with broad, multiwavelength, energy emission, as the case of the Crab Nebula, an emission model extracted from a single IACT does not cover the CTA observational window. The combination of different IACTs and other instruments (e.g. Fermi/LAT and AGILE/GRID for the lowest energies) and extrapolations in the missing energy bins are needed in the evaluation of many CTA science target spectral models.
\\
We consider three different spectral emissions that are commonly used in literature:
\begin{enumerate}
\item at first order, the Crab Nebula spectrum at VHE is expressed by a power-law model. We refer here to the C.U. definition used by [\citenum{2013APh....43..171B}] as sensitivity scale of the CTA Monte Carlo (MC) simulations:
\begin{equation}
\rm 1 C.U. = 2.79\times10^{-11}\;cm^{-2}s^{-1}TeV^{-1}\times \left(\dfrac{\rm E}{\rm 1 \;TeV}\right)^{-2.57}
\end{equation}
\item from the H.E.S.S. observation [\citenum{2006A&A...457..899A}], the energy spectrum of the Crab Nebula is found to follow a power-law with an exponential cut-off in the 0.5 - 30 TeV energy range. We extend this modelization to cover the CTA energy range: from 20 GeV to 300 TeV, using the following best fit parameters:
\begin{equation}
\rm 1 C.U. = 4.0\times10^{-11}\;cm^{-2}s^{-1}TeV^{-1}\times \left(\dfrac{\rm E}{\rm 1\; TeV}\right)^{-2.39}e^{\rm -E/(14.3\; \rm TeV)}
\end{equation}
\item from the cross-calibration of Fermi/LAT and IACTs (H.E.S.S [\citenum{2006A&A...457..899A}], MAGIC [\citenum{2011ICRC...12..147C}], HEGRA [\citenum{2000ApJ...539..317A}]) observations, the work of [\citenum{2010A&A...523A...2M}] obtains the spectral energy distribution of the Crab nebula from 1 GeV to 100 TeV. Since the Crab nebula flux is the sum of many models describing different portions of the energy range, we use the numerical modelization provided by [\citenum{2015arXiv150907408D}].
\end{enumerate}
The differential flux, expressed as $\rm E^{2}\times dN/dE$ in erg cm$^{-2}$ s$^{-1}$, for the three models are are plotted in Figure \ref{fig:model}.
All models agree within a factor 2 from $\sim200$ GeV to $\sim10$ TeV. Extending the H.E.S.S. observation at the CTA lowest and highest energies respectively overestimates and dramatically underestimates the flux obtained by [\citenum{2010A&A...523A...2M}].

\section{INTEGRAL SENSITIVITY FOR CRAB-LIKE SOURCES}
The sensitivity of a telescope generally refers to the minimum gamma-ray flux required to achieve a statistically significant detection of a scientific target. The flux is usually indicated as an amount of counts, or photons, or energy, for unit of area and time. 
The CTA sensitivity depends on (i) the CTA design (e.g. effective area, angular resolution), (ii) the reconstruction algorithms (e.g. the background rejection efficiency), (iii) the science target features (e.g. energy range, flux, variability), (iv) the statistical assumptions, defined by the characteristic amount of events under study.
The integral sensitivity at the energy E$_{0}$ is the minimum detectable gamma-ray flux at E$>$E$_{0}$. It integrates all the source photons from a minimum energy to the upper limit of CTA energy range.
Since each science target is characterized by a different spectral distribution, which affects the distribution of the photons in a given energy range, the integral sensitivity is not a general feature of the observatory but must be linked to the astrophysical object under study. 
\\
Summing all the source photons the information on the spectrum is lost, but the detection significance is enhanced. Given the short timescales on which RTA must operate (from minutes to hours) and the science alerting requirement, the integral sensitivity is the key parameter to define the RTA performance. For a preliminary evaluation of the RTA and Level-B differential sensitivity at exposures $\geq1000$ seconds see [\citenum{2015arXiv150901943F}].

\subsection{Methods and assumptions}
The computation presented here refers to the whole CTA Southern observatory, assumed to be composed of 100 telescopes and comprising the three telescopes sizes (4 LSTs, 24 MSTs, and 72 SSTs). 
\\
In a VHE observation, the detection of a source can be described in terms of an on/off problem: two observations are performed, one in an empty field (off) where there is no contribution from the source flux, and the other on the source (on). A third parameter $\alpha$ characterizes the observation. It is defined as the ratio of the different effective areas \textit{A}, exposure time \textit{t}, and size of the region \textit{k} for the two on/off observations on the sky:
\begin{equation}
\alpha = \frac{\rm k_{\rm on}\times\rm t_{\rm on}\times\rm A_{\rm on}}{\rm k_{\rm off}\times\rm t_{\rm off}\times\rm A_{\rm off}}
\end{equation}
Given N$_{\rm on}$ and N$_{\rm off}$ the number of counts for the on and off detections respectively, we define N$_{\rm s}$, the number of source counts, as N$_{\rm s}$ = N$_{\rm on}$ - $\alpha\rm N_{\rm off}$.
The significance of the minimum detectable flux depends on the required confidence level for detection. On the basis of the CTA standard rules [\citenum{2013APh....43..171B}], given N$_{\rm s}(>$E$_{0}$) the number of events above the energy E$_{0}$ from the source, we base our integral sensitivity evaluation on the following assumptions and techniques:
\begin{enumerate}
\item a statistical significance of 5 standard deviations above E$_{0}$ is required;
\item a minimum number of N$_{\rm s}$ = 10 must be collected for each detection;
\item an on to off exposure ratio of $\alpha=0.2$;
\item the excess above the background level must be greater than 5\%. 
\end{enumerate}
Each energy decade is divided in 5 energy bins with equal logarithmic bin width. For each lower limit of the energy bin, E$_{0}$, the number of background counts N$_{\rm off}$ for energies higher than E$_{0}$ is computed, and the minimum number of N$_{\rm s}(>$E$_{0}$) ensuring the statistical and significance assumptions is evaluated.
\\
Given the three parameters N$_{\rm on}$, N$_{\rm off}$, and $\alpha$, different statistical techniques allow us to assign a significance to the possible presence of a source in the on field.
At high energies and for long exposures, the significance S(N$_{\rm on}$, N$_{\rm off}$, $alpha$) is usally computed using equation (17) of [\citenum{1983ApJ...272..317L}]. 
  \begin{figure} [h!]
   \begin{center}
   \begin{tabular}{c} 
   \includegraphics[width=0.7\textwidth]{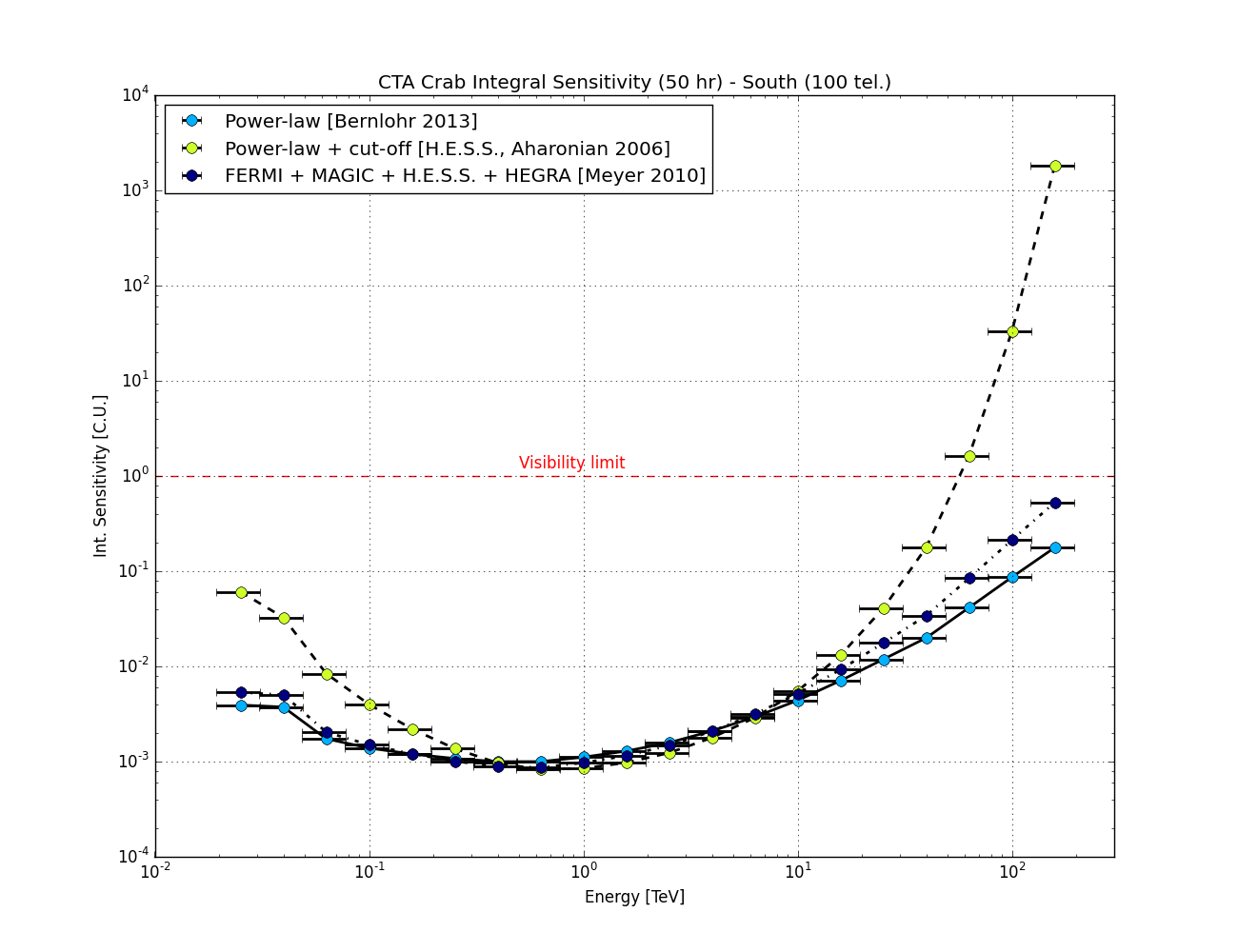}
   \end{tabular}
   \end{center}
   \caption 
   { \label{fig:nom}The CTA-South nominal Crab nebula integral sensitivity, for a standard 50 hour exposure, expressed in C.U. for the three emission models under analysis. The horizontal red line indicates 1 C.U. .}
   \end{figure}  
 \subsection{Impact of emission model on the nominal sensitivity}
The integral sensitivity, in C.U., resulting from a 50 hours observation of the Crab Nebula is plotted in Figure \ref{fig:nom} for the three emission models listed in Sec. \ref{sec:crab}, using the full CTA Southern observatory. 
An integral sensitivity of 1 mCrab is achieved above 1 TeV. When we refer to the model from [\citenum{2010A&A...523A...2M}] as the reference model for the Crab Nebula emission (black dash-dotted line of Fig. \ref{fig:nom}), a Crab-like source with 10\% of the Crab Nebula flux would be still visible at 70 TeV.

\subsection{On-Site Analysis integral sensitivity}
\begin{figure} [h!]
   \begin{center}
   \begin{tabular}{c} 
   \includegraphics[width=0.9\textwidth]{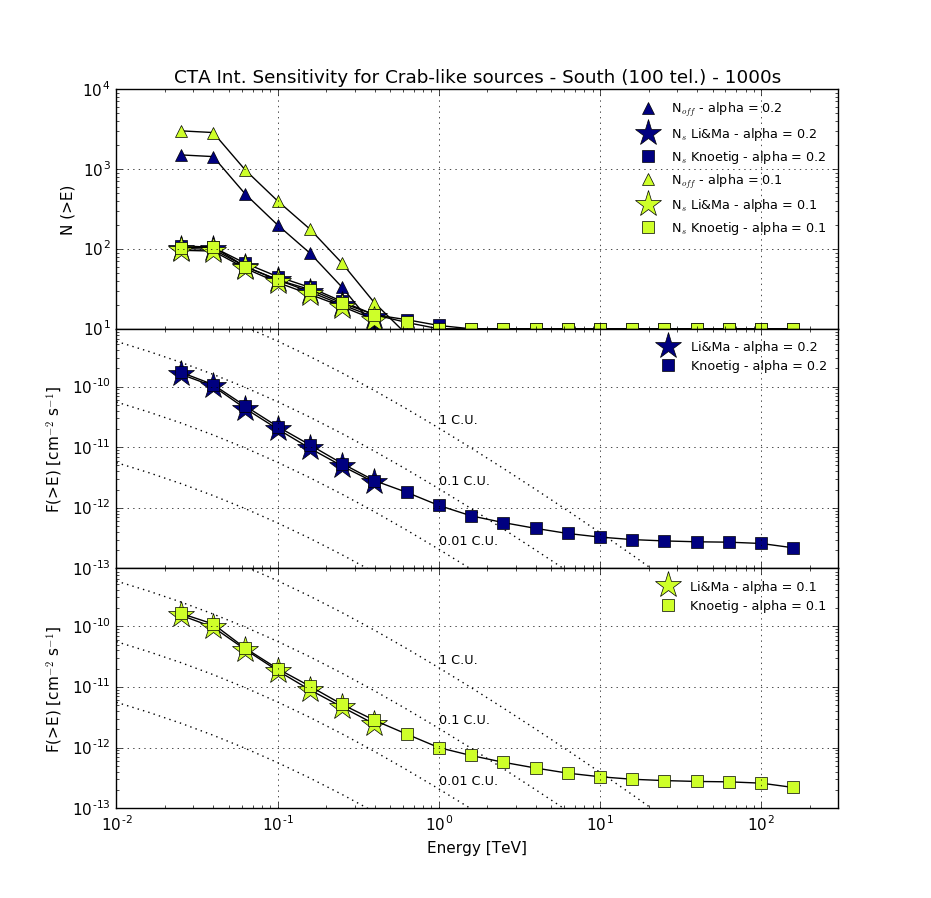}
   \end{tabular}
   \end{center}
   \caption 
   { \label{fig:bayes} \textit{Top panel}: the number of N$_{\rm off}$ (triangles) and N$_{\rm s}$  counts  as a function of energy for a simulated 1000 s exposure, using an on/off ratio $\alpha$ = 0.2 (default, in blue) and 0.1 (in yellow); \textit{middle and bottom panels}: the integral sensitivity for Crab-like sources for an exposure of 1000s , using an on/off ratio $\alpha$ = 0.2 (default, in blue) and 0.1 (in yellow). Stars and squares values are obtained using the significance computation from the Li\&Ma and the Knoetig procedure respectively.] 
}
   \end{figure}

\begin{figure} [h!]
   \begin{center}
   \begin{tabular}{c} 
   \includegraphics[width=0.8\textwidth]{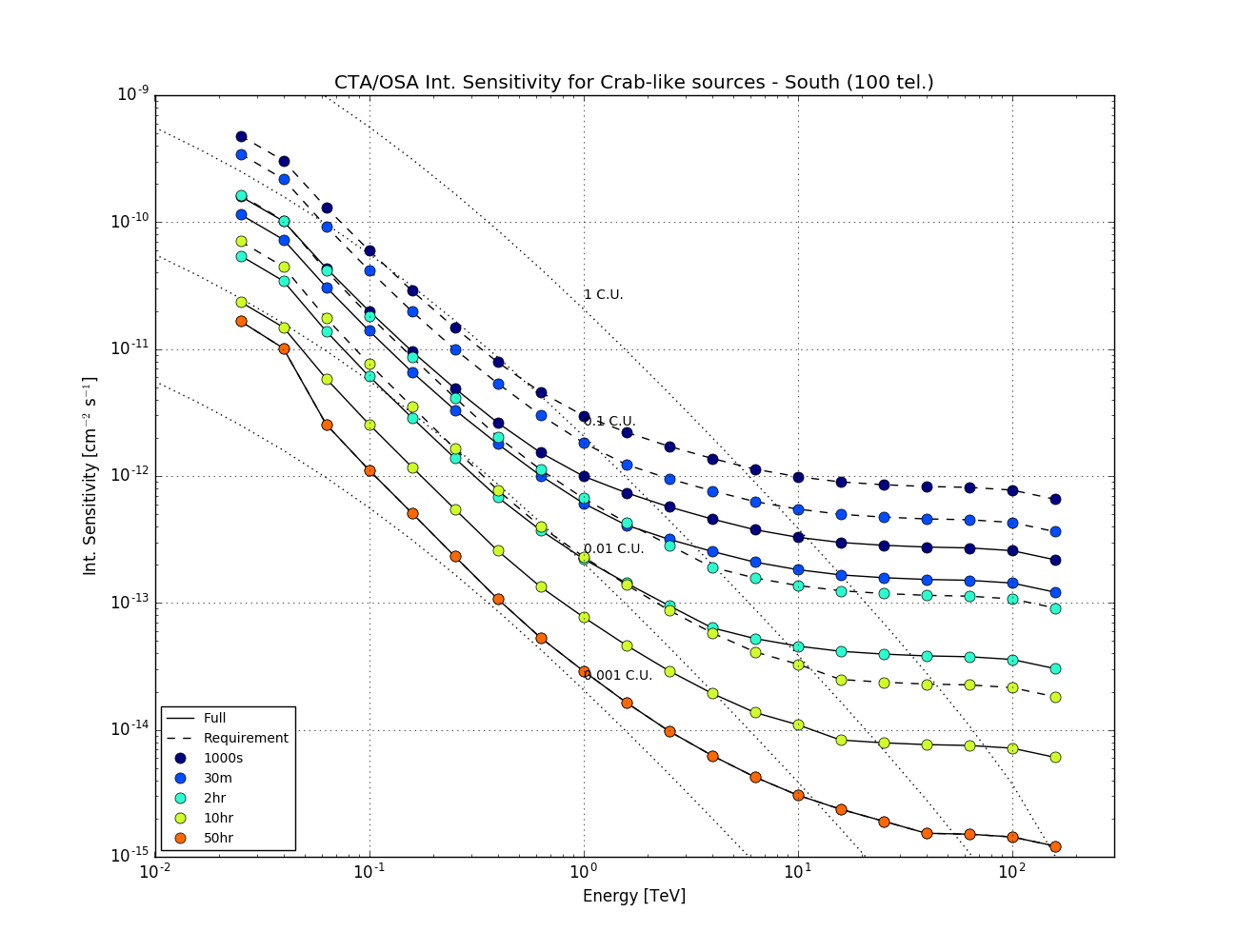}
   \end{tabular}
   \end{center}
   \caption 
   { \label{fig:osa}The CTA-South OSA and nominal Crab nebula integral sensitivity obtained for the following exposures: 1000 s, 30 minutes, 2 hours,  10 hours, and 50 hours. The continued and dashed line refer to the full and required (a factor 3 worse) sensitivity respectively. The whole southern array (100 telescopes) is used.
}
   \end{figure}

After selecting the model from [\citenum{2010A&A...523A...2M}] as the reference Crab Nebula emission model for the present analysis, the integral sensitivity is computed for short exposures, where the RTA and Level-B analysis will continuously monitor the transient VHE sky. The integral sensitivity evaluation presented here uses the effective area and the background rate from the PROD2 MC simulations, with the following exposures: 1000 seconds, 30 minutes, 2 hours, 10 hours, and 50 hours.
MC simulations for shorter exposures (second-minute timescales) with dedicated sub-array configurations are currently under development.
\\
The procedure described in [\citenum{1983ApJ...272..317L}] for the significance computation - referred here as Li\&Ma for clarity - is based on a likelihood ratio method and requires that the N$_{\rm on}$ and N$_{\rm off}$ values obtained from a single detection are \textit{not too few}. This assumption is not always true for very short exposures. As shown in Figure \ref{fig:bayes} (top panel), if a 1000s exposure is used the number of N$_{\rm off}$($>$E) events decreases below 10 if a minimum energy threshold above 400 GeV ($\alpha=0.2$) and 600 GeV ($\alpha=0.1$) is considered.
In the case of very low count rates, as the case of many VHE science cases and in particular all the ones involving short term transient phenomena, the assumption of a normally distributed likelihood distribution is not valid and the Li\&Ma can not be applied. The use of an objective Bayesian solution, using improper priors as a tool to produce proper posteriors, is proposed by [\citenum{2014ApJ...790..106K}, referred here as Knoetig] to compute the significance of an on/off detection without any limitation on the count number. While subjective Bayesian solutions lean on the knowledge of a prior opinion (e.g. a known background rate) to evaluate the probability of the posteriors, with Equation (23) of the objective Bayesian method we can rely on the N$_{\rm on}$, N$_{\rm off}$, and $\alpha$ values to assess the significance of a detection in spite of low count numbers. As in the Li\&Ma work, a null hypothesis probability of $5.7\times10^{-7}$ is equivalent to a 5$\sigma$ detection measurement.
\\
In Fig. \ref{fig:bayes} (middle and bottom panel) we compute the CTA integral sensitivity for a short 1000s exposure using the two methods to evaluate the significance. We assume a minimum value N$_{\rm off}\ge10$ as the requirement to apply the Li\&Ma procedure. For this reason, the Li\&Ma sensitivity curve stops at $\sim400$ GeV. In the validity range of the Li\&Ma procedure, the objective Bayesian and the likelihood ratio results are consistent and result in a similar minimum integral flux for CTA (the Knoetig method gives slightly lower significance values). At higher energies, where fewer background counts are detected, we only use the Knoetig procedure to evaluate the significance of the on/off detection. However, at higher energies the intrinsic low signal rate affects the detection and the minimum flux only depends on the ability to accumulate N$_{\rm s}\geq10$ source counts.
The top panel of Fig. \ref{fig:bayes} shows the N$_{\rm s}$ source count values, starting from the minimum threshold of 10, required to achieve a significance higher than 5$\sigma$: it is clear that above 400 GeV N$_{\rm s}$ reaches its minimum threshold and the resulting integral sensitivity curve is only affected by the effective area. Reducing by 50\% the on/off exposure ratio (from 0.2 to 0.1) has no significant impact on our results.
\\
\textbf{The Southern CTA array will be sensitive to at least 10\% of the integral Crab Nebula flux if a 1000s-long nominal pointing is executed and a lower energy threshold $<5$ TeV is used. An energy threshold of 10 TeV will still allow  to detect gamma-rays from the Crab Nebula in 1000s, thanks to the wide area covered by the MSTs and SSTs.}
\\
Figure \ref{fig:osa} compares the integral sensitivity for 1000 seconds, 30 minutes, 2 hours, 10 hours, and the standard 50 hour exposure.
Since the OSA pipeline could apply simplified algorithms and crude selection cuts for hadron/gamma separation to speed-up the analysis, the final sensitivity could be lower than the nominal, off-line, sensitivity. 
The CTA design requires a maximum integral sensitivity worsening factor of 3 for the RTA. Both cases, the full sensitivity (continuous line) and the conservative (dashed line) sensitivity evaluation, based on the CTA requirement, are plotted for exposures lower than 50 hours. The 50 hour nominal CTA sensitivity is plotted as reference.

\section{SUMMARY AND CONCLUSIONS}
The effect of different approximations for the emission model of the Crab Nebula on the CTA integral sensitivity is evaluated, with the aim of emphasizing the need for representative spectra of the CTA science targets in the evaluation of the OSA use cases. A full southern array integral sensitivity (50 hour exposure) of 1 mCrab is obtained for a minimum energy threshold lower than 1 TeV. The impact of an objective Bayesian method, instead of the standard likelihood ratio procedure, to compute the signal significance is tested throughout the CTA energy range for a 1000s exposure. In the Li\&Ma validity range ($<400$ GeV for a 1000s exposure), the two methods are consistent; at higher energies, i.e. at low N$_{\rm off}$ values, only the Bayesian method is used - even if the signal statistics limit dominates the sensitivity.
\\
Short exposures, from 1000 seconds to 10 hours, are used to evaluate the OSA and in particular the RTA performance in the detection of Crab-like sources: even for a 1000 second long observation, a significant Crab detection is ensured up to about 10 TeV.
We plan the computation of the integral sensitivity, using short exposures, for each RTA science use case [\citenum{2016SPIE}], using the most accurate models available from literature to characterize each science target class. 

\acknowledgments 
 
We gratefully acknowledge support from the agencies and organizations listed under Funding Agencies at this website: http://www.cta-observatory.org/. VF would like to thank Prof. E. Dolera (University of Pavia, Department of Mathematics) for fruitful discussions on statistics and significance and M. L. Ahnen (ETH Zurich) for carefully reading and commenting the paper. This paper has gone through internal review by the CTA Consortium.

\bibliography{fioretti_cta} 
\bibliographystyle{spiebib} 

\end{document}